\documentclass[twocolumn,showpacs,preprintnumbers,amsmath,amssymb]{revtex4}
\usepackage{epsfig,subfigure}
\usepackage{graphicx}
\usepackage{amsmath}
\usepackage{color}

\begin{document}

\preprint{FIS-UI-TH-04-03}

\title{Photoproduction of Pentaquark in Feynman and Regge Theories}

\author{T. Mart}
\affiliation{Departemen Fisika, FMIPA, Universitas Indonesia, Depok
16424, Indonesia}

\begin{abstract}
Photoproduction of the $\Theta^+$ pentaquark on the proton 
is analyzed by using an isobar and a Regge models.
The difference in the calculated total cross section is found to be more 
than two orders of magnitude for a hadronic form factor
cut-off $\Lambda > 1$ GeV. Comparable results would be obtained 
for $0.6\le \Lambda\le 0.8$ GeV. We also calculate contribution of the 
$\Theta^+$ photoproduction to the GDH integral. By comparing
with the current phenomenological calculation, it is found that
the GDH sum rule favors the result obtained from Regge approach
and isobar model with small $\Lambda$.
\end{abstract}


\maketitle

The observation of a narrow baryon state from the missing mass spectrum
of $K^+n$ and $K^+p$ with extracted mass $M=1540$ MeV  
\cite{leps,saphir,clas,diana,hermes} has led to a great 
excitement in hadronic and particle physics communities. 
This state is identified as the $\Theta^+$ pentaquark that has been previously
predicted in the chiral soliton model \cite{diakonov}. Since then
a great number of investigation on the $\Theta^+$ production has 
been carried out. In general, these efforts can be divided into two
categories, i.e., investigations using hadronic and electromagnetic processes.
The electromagnetic production (also known as the photon-induced
production) is, however, well known as a more ''clean'' process, 
since electromagnetic 
interaction has been well under-controlled. Furthermore, photoproduction
process provides an easier way to ''see'' the $\Theta^+$ which contains
an antiquark, since all required constituents are already present in
the initial state \cite{Karliner:2004gr}. Other processes, such as $e^+e^-$ and
${\bar p}p$ annihilations, would produce the strangeness-antistrangeness
(and baryon-antibaryon in the case of $e^+e^-$) from gluons, which has
a consequence of the suppressed cross section \cite{Titov:2004wt}.

Several $\Theta^+$ photoproduction studies have been performed by using 
isobar models with Born approximation 
\cite{ysoh,liu&ko2004,nam,qzhao,yu&ji2004,yrliu,wwli,pko}, with the
resulting cross section spans from several nanobarns to almost one $\mu$barn,
depending on the $\Theta^+$ width, parity, hadronic form factor cut-off, 
and the exchanged particles used in the process. Those parameters are 
unfortunately still uncertain at present.
Furthermore, the lack of information on coupling constants
has severely restricted the number of exchanged particles used in
the process, including a number of resonances which have been
shown to play important roles at $W$ around 2 GeV and determine the 
shape of cross sections of the $K\Lambda$ and $K\Sigma$ photoproductions
\cite{mart2000}.

Therefore, it is important to constrain the proliferation of models by
using all available informations in order to achieve a reliable
cross section prediction which is urgently required by present experiments.
For this purpose we will exploit the isobar and Regge models and using
all available coupling constant informations. The use of Regge model
has a great advantage since the number of uncertain parameters
is much less than those of the isobar one. Nevertheless,
from the experience in $K\Lambda$ and $K\Sigma$ photoproduction, 
in spite of using a small number of parameters Regge model works quite
well at high energies and the discrepancy with experimental data at 
the resonance region is found to be less than 50\% \cite{guidal97,Mart:2003yb}. 
Concerning with the high threshold energy of this process ($W\approx$ 
2 GeV) it is naturally imperative to consider the Regge mechanism
in the calculation. As an example, a reggeized isobar model has shown a great 
success in explaining experimental data of $\eta$ and $\eta '$ photoproductions
up to the photon lab energy $E_\gamma^{\rm lab}=2$ GeV \cite{Chiang:2002vq}.

\begin{figure*}[hbt]
\centering
 \mbox{\epsfig{file=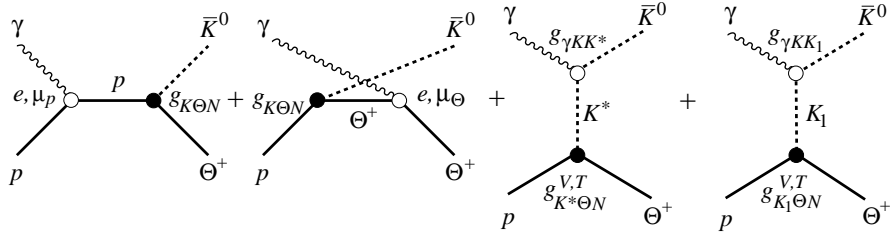,width=12cm}}
\caption{Feynman diagrams for $\Theta^+$ photoproduction on the proton 
  $\gamma + p \longrightarrow {\bar K}^0 + \Theta^+$.}\label{fig:feynman}
\end{figure*}

In this paper we compare the cross sections obtained from both models
and investigate the effect of hadronic form factor cut-off ($\Lambda$) variation 
in the isobar model. To this end, we will consider the positive parity of $\Theta^+$,
since previous calculations have found a ten times smaller cross section
if one used the negative parity state, whereas we concern very much with the 
overprediction of cross sections by isobar models. Moreover, our first 
motivation is to investigate the effect of $\Lambda$ variation and compare the 
varied cross sections with that of the Regge model. To further support our
finding, we will calculate contribution of the $\Theta^+$ photoproduction
to the Gerasimov-Drell-Hearn (GDH) integral from both models. Since only
GDH integral for proton is relatively well understood \cite{drechsel}, we 
calculate only photoproduction on the proton 
\begin{eqnarray*}
\gamma (k) + p (p) \longrightarrow {\bar K}^0 (q) + \Theta^+(p') ~.
\end{eqnarray*}
In isobar model the amplitudes 
are obtained from a series of tree-level Feynman diagrams shown in 
Fig.\,\ref{fig:feynman}. They contain the $p$, $\Theta^+$, $K^*$
and $K_1$ intermediate states. The neutral kaon $K^0$ cannot contribute 
to this process since a real photon cannot
interact with a neutral meson. The $K^*$ and $K_1$ intermediate states are 
considered here, since previous studies on $K\Lambda$ and $K\Sigma$ photoproductions
have proven their significant roles.
The transition matrix for both reactions can be decomposed into 
\begin{eqnarray}
M_{\mathrm fi} &=& {\bar u}({\mbox{\boldmath ${p}$}}') 
  \sum_{i=1}^{4} A_i~M_i ~u({\mbox{\boldmath ${p}$}}) ~,
\label{eq:mfi}
\end{eqnarray}
where the gauge- and Lorentz invariant matrices $M_i$ are given 
in, e.g., Ref.\,\cite{Lee:1999kd}.
In terms of the Mandelstam variables $s$, $u$, and $t$, 
the functions $A_i$ are given by
\begin{eqnarray}
\label{eq:a1}
A_{1} &\!\! =\! & -\frac{e g_{K \Theta N} F_1(s) }{s - m_{p}^{2}} \left(1 +
\kappa_{p} \frac{m_{p} - m_{\Theta}}{2 m_{p}} \right)\nonumber\\
&&  - \frac{e g_{K \Theta N} F_2(u)}{u - m_{\Theta}^{2} + im_\Theta\Gamma_\Theta} \!
 \left[ 1 + \frac{\kappa_{\Theta}\left(
m_{\Theta} - m_{p} - {\textstyle \frac{i}{2}}\Gamma_\Theta\right)}{2 m_{\Theta}}  
\right] \nonumber\\
&& - \frac{G^TF_3(t)}{M(t-m_{K^*}^2+im_{K^*}\Gamma_{K^*})(m_\Theta + m_p)} ~,\\
A_{2} & = & \frac{2e g_{K \Theta N}}{t - m_{K}^{2}} 
\left(\frac{1}{s - m_{p}^{2}} + \frac{1}{u - m_{\Theta}^{2}} \right)
{\widetilde F}(s,u,t) \nonumber\\&& 
+ \frac{G^T_{K^*}F_3(t)}{M(t-m_{K^*}^2+im_{K^*}\Gamma_{K^*})(m_\Theta + m_p)}
\nonumber\\&&
- \frac{G^T_{K_1}F_3(t)}{M(t-m_{K_1}^2+im_{K_1}\Gamma_{K_1})(m_\Theta + m_p)}
,\label{eq:ftilde}\\
A_{3} & = & \frac{e g_{K \Theta N}}{s - m_{p}^{2}}~
\frac{\kappa_{p} F_1(s)}{2 m_{p}} - \frac{e g_{K \Theta N}}{u - 
m_{\Theta}^{2}}~ \frac{\kappa_{\Theta} F_2(u)}{2 m_{\Theta}} \nonumber\\&& 
- \frac{G^T_{K^*}F_3(t)}{M(t-m_{K^*}^2+im_{K^*}\Gamma_{K^*})}
\frac{m_\Theta - m_p}{m_\Theta + m_p}\nonumber\\&& 
+ \frac{(m_\Theta +m_p)G^V_{K_1}+ 
(m_\Theta -m_p)G^T_{K_{1}}}{M(t-m_{K_1}^2+im_{K_1}\Gamma_{K_1})}
\frac{F_3(t)}{m_\Theta + m_p} ,\nonumber\\ \\
A_{4} & = & \frac{e g_{K \Theta N}}{s - m_{p}^{2}}~
\frac{\kappa_{p} F_1(s)}{2 m_{p}} +
\frac{e g_{K \Theta N}}{u - m_{\Theta}^{2}}~ 
\frac{\kappa_{\Theta} F_2(u)}{2 m_{\Theta}} \nonumber\\&& 
+ \frac{G^V_{K^*}F_3(t)}{M(t-m_{K^*}^2+im_{K^*}\Gamma_{K^*})}~ ,
\label{eq:a4}
\end{eqnarray}
with $\kappa_p$ and $\kappa_\Theta$
indicate the anomalous magnetic moments of the proton and $\Theta$, and $M$
is taken to be 1 GeV in order to make the coupling constants
\begin{eqnarray}
  \label{eq:gv_gt}
  G^{V,T}_{K^*(K_1)\Theta N} &=& g^{V,T}_{K^*(K_1)\Theta N}\, g_{K^* K\gamma}
\end{eqnarray}
dimensionless.

The inclusion of hadronic form factors at hadronic vertices is performed 
by utilizing the Haberzettl prescription \cite{{Haberzettl:1998eq}}. 
The form factors in this calculation are taken as
\begin{eqnarray}
  \label{eq:form_factor}
  F_i(q^2) &=& \frac{\Lambda^4}{\Lambda^4 + (q^2-m_i^2)^2} 
\end{eqnarray}
with $q^2=s,u,t$, and $i=p,\Theta,{\bar K}$, while $\Lambda$ the corresponding cut-off.
The form factor for non-gauge-invariant terms ${\widetilde F}(s,u,t)$ 
in Eq.\,(\ref{eq:ftilde}) is extra constructed in order to satisfy crossing 
symmetry and to avoid a pole in the amplitude 
\cite{Davidson:2001rk}.

The coupling constant $g_{K\Theta N}$ is calculated from the decay width of
the $\Theta^+\to K^+ n$ by using
\begin{eqnarray}
  \label{eq:width}
  \Gamma &=& \frac{g^2_{K^- \Theta^+ n}}{4\pi}\,\frac{E_n-m_n}{m_\Theta}\, p ~,
\end{eqnarray}
with
\begin{equation}
  p = \frac{[\{m_\Theta^2-(m_K+m_n)^2\}\{m_\Theta^2-(m_n-m_K)^2\}]^{1/2}}{2m_\Theta}
\end{equation}
The precise measurement of the decay width is still lacking due to the
experimental resolution. The reported width is in the range of 6--25 MeV
\cite{leps,saphir,clas,diana,hermes,svd,nusinov}. Theoretical analyses of 
 $K^+N$ data result in $\Gamma\le 1 $ MeV
 \cite{theor_anal}, whereas the Particle Data Group \cite{pdg2004,cahn} announces 
$\Gamma = 0.9\pm 0.3 $ MeV. Based on this information, we decided to use a width of
1 MeV in our calculation. We find that the isobar model becomes no longer 
sensitive to the value of $g_{K\Theta N}$ coupling constant, once we have included
the $K^*$ and $K_1$ exchanges. Explicitly, we use 
\begin{eqnarray}
  \label{eq:cc_num}
  \frac{g_{K\Theta N}}{\sqrt{4\pi}} &=& 0.39 ~.
\end{eqnarray}
The magnetic moment of $\Theta^+$ is also not well known. A recent chiral soliton 
calculation \cite{kim2003} yields a value of $\mu_\Theta = 0.82 ~ \mu_N$,
from which we obtain $\kappa_\Theta=0.35$. As in the case of $g_{K\Theta N}$ 
coupling constant, our calculation is also not sensitive to the numerical
value of $\Theta^+$ magnetic moment, so that we feel it is save to use the above value.
Note that the Regge model does not depend on this coupling constant as well as
the $\Theta^+$ magnetic moment.

The transition moment is related to the radiative decay width by 
\begin{eqnarray}
  \label{eq:transition_moment}
  \Gamma_{K^{*}\to K\gamma} = \frac{\alpha}{24} 
  \left(\frac{g_{K^{*} K\gamma}}{M}\right)^2\, \left[m_{K^*}\left(1-
    \frac{m_{K}^2}{m_{K^*}^2}\right)\right]^3 .
\end{eqnarray}
The decay width for $K^{*0}(892)$ is well known, i.e. \cite{pdg2004}
\begin{eqnarray}
\Gamma_{K^{*0}\rightarrow K^0\gamma} &=& 116\pm 10 ~~{\rm keV} ~.
\end{eqnarray}
Thus, we obtain $g_{K^{*0}K^0\gamma}=-1.27$, 
where we have used the quark model prediction of Singer and Miller 
\cite{singer} in order to constrain the relative sign.

The coupling constants $g^{V}_{K^{*} \Theta N}$ and 
$g^{T}_{K^{*} \Theta N}$ are also not well known. Therefore, we follow 
Refs.\,\cite{yu&ji2004,liu&ko2004}, i.e., using $g^{V}_{K^{*} \Theta N}=1.32$
and neglecting $g^{T}_{K^{*} \Theta N}$ due to lack of information
on this coupling. By combining the electromagnetic and hadronic coupling
constants we obtain
\begin{eqnarray}
  \label{eq:GVK*}
  \frac{G^{V}_{K^*\Theta N}}{4\pi} &=& 8.72\times 10^{-2} ~.
\end{eqnarray}

Most previous calculations excluded the $K_1$ exchange, mainly due to the
lack of information on the corresponding coupling constants. Reference 
\cite{yu&ji2004} used the vector dominance relation 
$g_{K_1K\gamma}=eg_{K_1K\rho}/f_\rho$ to determine the electromagnetic
coupling $g_{K_1K\gamma}$, where $f^2_\rho/4\pi=2.9$ and
$g_{K_1K\rho}=12$ is taken from the effective Lagrangian calculation
of Ref.\,\cite{haglin94}. As in the case of $K^*$, the $K_1$ hadronic 
tensor coupling will be neglected in this calculation due to the same 
reason. Following Ref.\,\cite{yu&ji2004}, the $K_1$ axial vector 
coupling $g^{V}_{K_1\Theta N}$ is estimated from an isobar model
for $K^+\Lambda$ photoproduction by using the extracted 
$G^{V}_{K^*\Lambda N}/ G^{V}_{K_1\Lambda N}$ ratio. However,
instead of using the result of WJC model \cite{wjc} we will
exploit the extracted ratio found in Ref.\,\cite{mart2000}.
There are two models given in Ref.\,\cite{mart2000}, i.e., models
with and without the missing resonance $D_{13}(1895)$, which give
a ratio of $-0.24$ and $-8.25$, respectively. Incidentally, 
Ref.\,\cite{wjc} gives a ratio of $-8.26$, i.e., similar to
the model without missing resonance. In our calculation we will
use this ratio and excluding the result from the model with 
missing resonance, since the later leads to a divergence
contribution to the GDH sum rule, as
will be described later. In summary, in our calculation we use
\begin{eqnarray}
  \label{eq:GVK1}
  \frac{G^{V}_{K_{1}\Theta N}}{4\pi} &=& -7.64\times 10^{-3} ~.
\end{eqnarray}
The cross section can be easily calculated from the functions $A_i$ given by
Eqs.\,(\ref{eq:a1})--(\ref{eq:a4}) \cite{deo}. 

For the Regge model one should only use the last two diagrams in 
Fig.\,\ref{fig:feynman}.
Hence, the result from Regge model will not depend on the value
of $g_{K\Theta N}$ and $\Theta^+$ magnetic moment.
The procedure is adopted from Ref.\,\cite{guidal97}, i.e., by 
replacing the Feynman propagator with the Regge propagator
\begin{eqnarray}
  \label{eq:regge}
  P_{\rm Regge} &=& \frac{s^{\alpha_{K^i}(t)-1}}{\sin [\pi\alpha_{K^i}(t)]}
                    ~ e^{-i\pi\alpha_{K^i}(t)} ~ 
                    \frac{\pi\alpha_{K^i}'}{\Gamma [\pi\alpha_{K^i}(t)]} ~,
\end{eqnarray}
where $K^i$ refers to $K^*$ and $K_1$, and 
$\alpha_{K^i} (t) = \alpha_0 + \alpha '\, t$ denotes the corresponding
trajectory \cite{guidal97}. We note that Ref.\,\cite{guidal97} used
form factors for extending the model to larger momentum transfer
(``hard'' process region). In our calculation we do not use these
form factors, since the corresponding cross sections 
at this region are already quite small and, therefore, will not strongly
influence the result of our calculation. We also note that systematic
analyses of experimental data on $\rho$, $\omega$, and $J/\Psi$ 
photoproductions explicitely require hadronic form factors
\cite{regge_rho}.

In both models, however, we can also calculate the spin 
dependent total cross sections 
\begin{eqnarray}
  \label{total_spin_dependent}
  \sigma_{\rm T} = \frac{\sigma_{3/2} + \sigma_{1/2}}{2} ~~~\textrm{and}~~
  \sigma_{\rm TT'} = \frac{\sigma_{3/2} - \sigma_{1/2}}{2},
\end{eqnarray}
where the latter is of special interest since it can be related to the 
proton anomalous magnetic moment $\kappa_p$ using the GDH sum rule
 \begin{equation}
  \label{gdh-sum-rule}
  -\frac{2\pi^2\alpha\kappa_p^2}{m_p^2} = \int_0^\infty 
  \frac{d\nu}{\nu} [ \sigma_{1/2}(\nu)-\sigma_{3/2}(\nu)]
  ~\equiv~ I_{\rm GDH} ,
\end{equation}
with $\nu=E_{\gamma}^{\rm lab}$ and $\sigma_{1/2}$ ($\sigma_{3/2}$) represents 
the cross section for possible proton and photon spin combinations  
with a total spin of $1/2$ ($3/2$). Thus,
we can calculate contribution of the $\Theta^+$ photoproduction
to the GDH integral $I_{\rm GDH}$ defined by Eq.\,(\ref{gdh-sum-rule}).
Note that in deriving Eq.\,(\ref{gdh-sum-rule}) it has been assumed that 
the scattering amplitude goes to zero in the limit of $|\nu| \rightarrow \infty$
\cite{sbass}.
\begin{figure}[t]
\centering
 \mbox{\epsfig{file=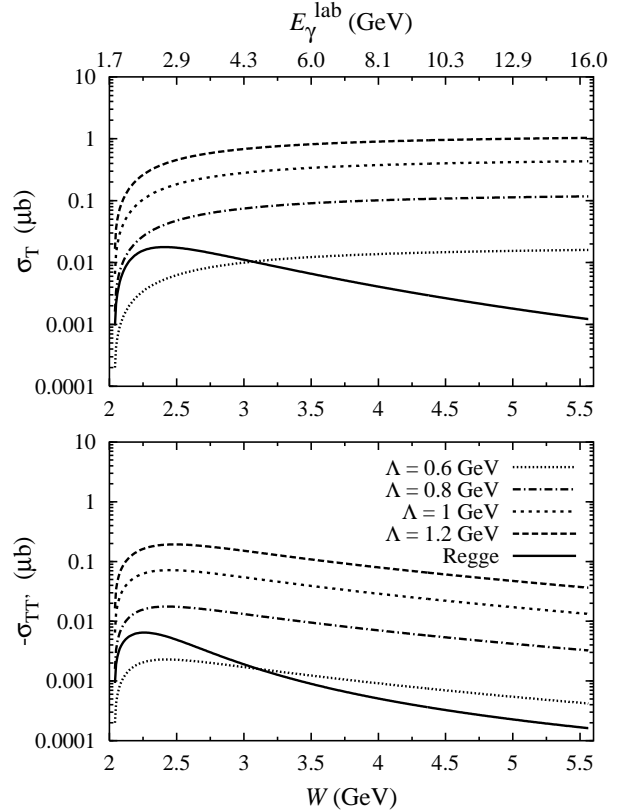,width=8.5cm}}
\caption{Total cross sections $\sigma_{\rm T}$ and $-\sigma_{\rm TT'}$
  of the isobar and Regge models. In isobar model variation of the
  total cross section for different hadronic form factor cut-offs is
  shown.}\label{fig:st}
\end{figure}   

The result of our calculation is depicted in Fig.\,\ref{fig:st},
where we compare the total cross section obtained from isobar
model with different hadronic cut-offs and that from the Regge model.
Obviously, the hadronic cut-off strongly controls
the magnitude of the cross section in the isobar model. By varying
$\Lambda$ from 0.6 to 1.2, both total cross sections increase
by two orders of magnitude, whereas their shapes 
remain stable and tend to saturate at high energies.
In the Regge model, both $\sigma_{\rm T}$ and $-\sigma_{\rm TT'}$
steeply rise to maximum at $W$ around 2.2 GeV and monotonically
decrease after that. Regge cross sections are clearly more convergent
than isobar ones. From threshold up to $W=3$ GeV, 
the cross section magnitude of the Regge model falls between 
the results obtained from isobar model with $\Lambda =0.6$ and
0.8 GeV. Starting from $W=3$ GeV, the magnitude becomes smaller
than the result from isobar model with  $\Lambda =0.6$ GeV.
Thus, future calculation should consider the hadronic cut-off 
in the range of 0.6 and 0.8 GeV.

\begin{figure}[t]
\centering
 \mbox{\epsfig{file=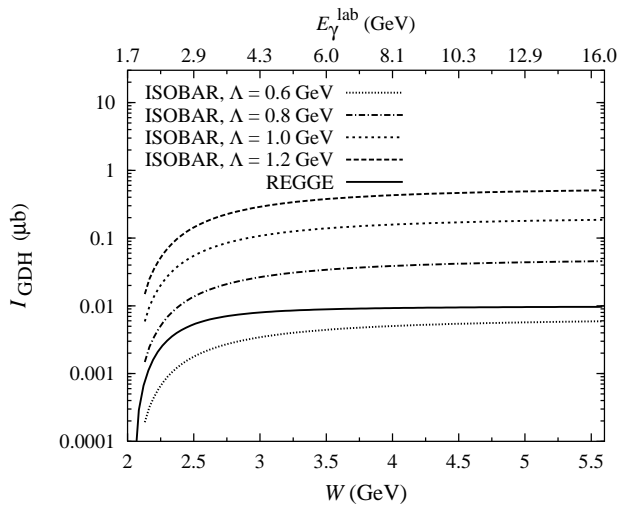,width=8.5cm}}
\caption{Contribution of the $\Theta^+$ photoproduction to the
  GDH integral of the proton for isobar (with different hadronic
  form factor cut-offs) and Regge models.}\label{fig:igdh}
\end{figure}   

Contribution from the pentaquark photoproduction to the GDH integral 
is shown in Fig.\,\ref{fig:igdh}, where we compare the result from
isobar and Regge models as in Fig.\,\ref{fig:st}. Clearly, the 
contribution is positive and small [note that direct calculation of the l.h.s.
of Eq.\,(\ref{gdh-sum-rule}) gives $-205~\mu$b]. Nevertheless, the
positive contribution to $I_{\rm GDH}$ invites an interesting discussion
if we consider the current knowledge of the GDH individual contribution
on the proton. By summing up contributions from $\pi$, $\eta$, $\pi\pi$,
and $K$ photoproduction, including contribution from the higher energy part, 
Ref.\,\cite{drechsel} found an $I_{\rm GDH}=-202~\mu$b. Recent calculation 
on vector meson ($\omega$, $\rho^0$, and $\rho^+$) contributions \cite{qzhao} 
indicates that their total contribution is also small ($+0.26~\mu$b).
From this point of view, negative (or positive but small) contribution
is more preferred. In other words, prediction from Regge model is
more desired rather than those of isobar model with $\Lambda \ge 0.8$ GeV.

As previously mentioned, the isobar model which includes the missing resonance 
\cite{mart2000} yields a ratio of 
$G^{V}_{K^*\Lambda N}/ G^{V}_{K_1\Lambda N}=-0.24$. Using
this ratio, we found that the predicted $-\sigma_{\rm TT'}$ flips to negative values
at $W$ around 3 GeV and starts to diverge from that point. 
This behavior merely emphasizes that certain mechanism (resonance
exchanges) is missing in the process. Therefore, in our calculation
we do not use this ratio.

\begin{figure}[t]
\centering
 \mbox{\epsfig{file=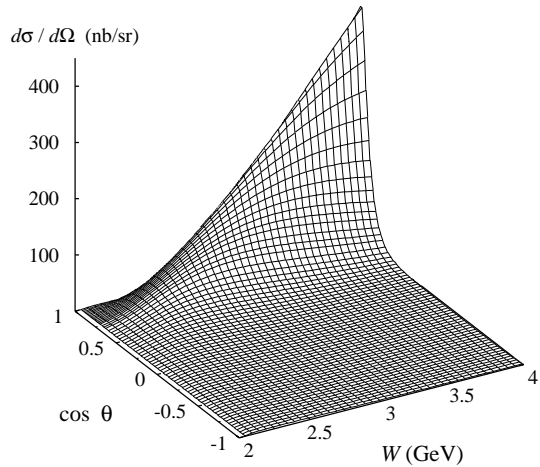,width=8.5cm}}
\caption{Differential cross section for $\Theta^+$ photoproduction 
  on the proton as functions of $\cos\theta$ and $W$ from isobar 
  model obtained with $\Lambda =1$ GeV. The same pattern,
  but with the magnitude 20 times smaller, would be obtained if
  one used $\Lambda =0.6$ GeV.}\label{fig:diffcs}
\end{figure}   

Recent isobar calculation for $K^+\Lambda$ photoproduction \cite{janssen}
claimed that a soft hadronic form factor (small $\Lambda$) is not
desired by the field theory. A harder form factor is achieved by 
including some $u$-channel resonances in the model. 
However, the authors do not build an 
explicit relation of this statement with the field theory. 
At tree level the extracted coupling constants are assumed 
to effectively absorb some important ingredients in the process, such
as rescattering terms and higher order corrections, which are
clearly beyond the scope of an isobar model. Therefore, the constants cannot 
be separated from the form factors. Together, they
define the effective coupling constants. Hence, it is hard to say
that at tree level calculation an isobar model should simultaneously produce 
SU(3) coupling constants and large cut-offs, i.e., weak suppression
on the divergent Born terms. A careful examination on the $u$-channel
resonance coupling constants reveals the fact that the corresponding
error bars are relatively large, which indicates that the inclusion
of these resonances is trivial \cite{petr}.

\begin{figure}[t]
\centering
 \mbox{\epsfig{file=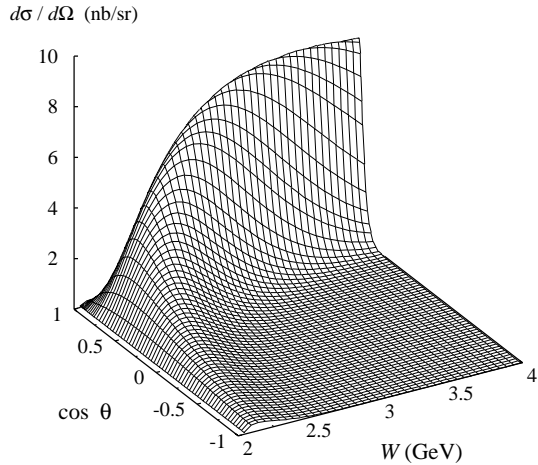,width=8.5cm}}
\caption{Differential cross section for $\Theta^+$ photoproduction on the
  proton as functions of $\cos\theta$ and $W$ from Regge model.}\label{fig:diffcs_reg}
\end{figure}   

The predicted differential cross sections are shown in Figs.\,\ref{fig:diffcs}
and \ref{fig:diffcs_reg}. The result shown in Fig.\,\ref{fig:diffcs} is obtained by
using $\Lambda =1$ GeV. By varying  the $\Lambda$ value, only the magnitude of
the cross section changes, whereas its shape with respect to $W$ and $\cos\theta$
remains stable. Thus, the difference between isobar and Regge models is quite
apparent in these figures. The isobar model limits measurements only at 
$0\le\cos\theta\le 0.5$, while Regge model allows for a complete angular
distribution of differential cross section at energies between
threshold and 2.5 GeV. At smaller $\cos\theta$ the cross section increases
with $W$ and becomes constant for $W>3.5$ GeV, in contrast to the prediction
from isobar model, which sharply increases as a function of $W$.
Future experimental measurements at JLab, SPRING-8, or 
ELSA will certainly be able to settle this problem.

In conclusion we have  simultaneously investigated $\Theta^+$ photoproduction
by using isobar and Regge models. We found that 
a comparable result is achieved if we use a hadronic cut-off 
between 0.6--0.8 GeV. This result indicates that previous
calculations which used a harder hadronic form factor are 
probably overestimates. By calculating the contribution
to the GDH integral we found that Regge model and isobar model
with $\Lambda \le 0.6$ GeV are favorable.

This work has been supported in part by the QUE (Quality for Undergraduate 
Education) project.

\end{document}